\documentclass[final,3p,times]{elsarticle}

\usepackage{graphics,epsfig,subfigure}
\usepackage{hyperref}

\usepackage{amsmath}
\usepackage{amssymb}
\usepackage{cases}

\usepackage{amstext}
\usepackage{color}
\usepackage{array}
\usepackage{multirow}
\usepackage{slashed}

\def\bal#1\eal{\begin{align}#1\end{align}}
\newcommand\beq{\begin{equation}}
\newcommand\eeq{\end{equation}}
\newcommand\beqn{\begin{eqnarray}}
\newcommand\eeqn{\end{eqnarray}}
\newcommand\nn{\nonumber}
\newcommand\fc{\frac}
\newcommand\lt{\left}
\newcommand\rt{\right}
\newcommand\pt{\partial}
\newcommand\tx{\text}
\newcommand\mc{\mathcal}
\allowdisplaybreaks

\biboptions{comma,square,compress,numbers,sort}

\journal{Physics Letters B}

\begin{document}

\begin{frontmatter}



\title{Linear Perturbations and Stability Analysis in $f(T)$ Braneworld Scenario}


\author[]{Ju-Ying Zhao}
\author[]{Mao-Jiang Liu}
\author[]{Ke Yang\corref{cor1}}
  \cortext[cor1]{keyang@swu.edu.cn, corresponding author} 

\address[]{School of Physical Science and Technology, Southwest University, Chongqing 400715, China}

\begin{abstract}

We conduct a detailed analysis of the full linear perturbations in the braneworld scenario within $f(T)$ gravity. By decomposing the perturbations of the theory into the scalar, transverse vector, antisymmetric pseudotensor, and symmetric transverse-traceless tensor modes, we derive the quadratic action for each mode. The results indicate that there is a total of one scalar, one massless vector, and one tensor propagating degrees of freedom. Consequently, in comparison to general relativity, no additional degrees of freedom appear under the flat braneworld background in the linearized theory. For a thick brane model with $f(T)=T+ \alpha T^2$, we find that it exhibits stability against linear perturbations.

\end{abstract}


\end{frontmatter}



\section{Introduction}

Although daily experience suggests that our spacetime is four-dimensional (4D), research into the existence of extra dimensions dates back to the 1920s, when the renowned Kaluza-Klein (KK) theory was proposed \cite{Kaluza1921,Klein1926}. The theory extends general relativity (GR) into five-dimensional (5D) spacetime by incorporating a compactified extra dimension at the Planck length. In KK theory, electromagnetism with $U(1)$ gauge symmetry arises naturally from the rotational symmetry along the compactified extra dimension \cite{Overduin1997}. Thus, this framework offers deeper insights into fundamental forces and the structure of  spacetime. 

Unlike KK theory, where matter fields are distributed throughout the entire higher-dimensional spacetime, the braneworld scenario suggests that our universe exists on a 3-brane within a higher-dimensional bulk, where all Standard Model particles are confined to the brane, while gravity is free to propagate into the extra dimensions \cite{Rubakov2001,Csaki2004}. This framework offers potential solutions to several longstanding problems in particle physics and cosmology, such as the gauge hierarchy problem \cite{Arkani-Hamed1998, Randall1999}, cosmic acceleration \cite{Deffayet2002}, and the cosmological constant problem \cite{Arvanitaki2017}. 

One of the prominent braneworld models is the ADD model proposed in 1998 \cite{Arkani-Hamed1998, Antoniadis1998}. It suggests a mechanism whereby gravity propagates into large extra dimensions, resulting in its dilution and weakening. Consequently, this provides a natural explanation for the gauge hierarchy problem. Later, Randall and Sundrum introduced the RS1 model \cite{Randall1999}, which consists of two 3-branes at the boundaries of a 5D anti-de Sitter bulk. This model generates a large hierarchy between the Planck scale and the electroweak scale without requiring extra dimensions that are significantly larger than the Planck length. Notably, as the size of the extra dimension approaches infinity, the RS1 model transitions into the RS2 model, which contains a single brane where the gravity remains effectively 4D \cite{Randall1999a}.

The thick brane scenario is considered a smooth generalization of the RS2 model \cite{DeWolfe2000,Gremm2000a,Csaki2000a}. In order to recover a proper 4D effective theory on the brane, it is essential for the massless modes of bulk matter fields to be localized there. The localization mechanisms and properties of various matter fields, including gauge bosons \cite{Chumbes2012,Cruz2013,Zhao2014,Alencar2014,Vaquera-Araujo2015,Zhao2015} and Dirac fermions \cite{Ringeval2002,Koley2005a,Melfo2006,Liu2008a,Liu2009b,Liu2013a,Barbosa-Cendejas2015}, have been intensively investigated in thick brane models. In addition to GR, researchers have proposed numerous modified gravity theories to address challenging cosmological issues such as singularities, dark energy, and dark matter. Thus, the thick brane scenario has also been generalized within the frameworks of modified gravities. For a  review and further references, see Refs.~\cite{Dzhunushaliev2010a,Liu2018}.

Among the various modified gravity theories, $f(T)$ gravity has gained considerable attention, see \cite{Cai2016,Bahamonde2023} for a comprehensive review. In $f(T)$ gravity, the fundamental dynamic field is the vielbein (or vierbein/tetrad in 4D spacetime), and the underlying spacetime is a curvature-free Weitzenb\"ock spacetime, rather than a torsion-free Riemannian one. This theory provides potential solutions to various challenges in cosmology, such as inflation \cite{Ferraro2007,Ferraro2008}, cosmic acceleration \cite{Bengochea2009,Linder2010}, and the Big Bang singularity \cite{Cai2011}. 

In Ref.~\cite{Yang2012}, Yang et al.~developed thick brane models within $f(T)$ gravity using a polynomial form, $f(T)=T+\alpha T^n$. Later, Menezes studied thick brane models with bulk scalar fields, which include both standard and generalized dynamics \cite{Menezes2014}. Wang et al.~constructed thick brane models incorporating various specific forms of $f(T)$ gravity with a noncanonical bulk K-field \cite{Wang2018a}. In Ref.~\cite{Yang2018}, a thick brane model was constructed in a reduced Born-Infeld-$f(T)$ theory. In Ref.~\cite{Guo2020a}, thick branes in mimetic $f(T)$ gravity and graviton resonances were investigated. The linear tensor perturbation analysis for $f(T)$ braneworld scenario was first conducted by Guo et al.~in Ref.~\cite{Guo2016}. Later, gravitational resonances in various $f(T)$-brane models were investigated in Ref.~\cite{Tan2021}. String-like braneworlds in $f(T)$ gravity were reported in Refs.~\cite{Moreira2021,Moreira2021a}. Additionally, branewold models in other gravitational theories within Weitzenb\"ock spacetime were studied in Refs.~\cite{Behboodi2013,Behboodi2014,Geng2014a,Behboodi2015,Moreira2021c,Moreira2021b,Yang2022a,Moreira2023a}.

In braneworld scenario, perturbation analysis is crucial for understanding the stability and dynamics of brane configurations in higher-dimensional spacetimes. Since the 5D local Lorentz invariance is broken in $f(T)$ gravity, the broken gauge freedom in the tangent frame results in 10 additional degrees of freedom (DOFs) in the vielbein. Consequently, the perturbation analysis in $f(T)$ theories is more complex than in GR. Nonetheless, apart from the tensor perturbation \cite{Guo2016}, other linear perturbations remain unexamined in $f(T)$ braneworld scenario. Therefore, in this work, we will analyze the full linear perturbations in $f(T)$ braneworlds.

The layout of the paper is as follows: In Sec.~\ref{Sec_Braneworld}, we briefly introduce the braneworld scenario in $f(T)$ gravity. In Sec.~\ref{Sec_Perturbation}, we analyze the full linear perturbations and their stabilities in $f(T)$ braneworlds. Finally, brief conclusions are presented in Sec.~\ref{Conclusions}. Throughout the work, the capital Latin indices $A,B,\cdots$ and $M,N,\cdots$ denote the 5D coordinates of tangent space and spacetime, respectively. In contrast, the small Latin indices $a,b,\cdots$ and Greek indices $\mu,\nu,\cdots$ represent the 4D coordinates of tangent space and spacetime, respectively.

\section{Braneworld scenario in $f(T)$ gravity}\label{Sec_Braneworld}

In $f(T)$ gravity, the fundamental dynamic field is the vielbein ${e^{A}}_{M}$. The metric tensor can be constructed from the vielbein through the relation $g_{MN}={e^{A}}_{M}{e^{B}}_{N}\eta_{AB}$, with $\eta_{AB} = \text{diag}(-1, 1, \cdots,1)$ the Minkowski metric for the tangent space. The Weitzenb\"ock connection is then defined as ${\Gamma^{P}}_{MN}={e_{A}}^{P}\pt_{N}{e^{A}}_{M}$. The torsion tensor, which describes the twisting of a space when moving along different paths, is built by the antisymmetric part of the Weitzenb\"ock connection, i.e., ${T^{P}}_{MN}={\Gamma^{P}}_{[NM]}$. The contorsion tensor is defined as the difference between the Weitzenb\"ock connection ${\Gamma^{P}}_{MN}$ and the Levi-Civita connection $\{{}^{P}{}_{MN}\}$, i.e.,  
\bal
{K^{P}}_{MN}&\equiv{\Gamma^{P}}_{MN}-\{{}^{P}{}_{MN}\}=\fc{1}{2}({{T_{M}}^{P}}_{N}+{{T_{N}}^{P}}_{M}-{{T^{P}}_{M}}_{N}).
\eal
Furthermore, with the torsion tensor and contorsion tensor, the superpotential torsion tensor is given by
\bal
{S_P}^{MN}=\fc{1}{2}\lt(K^{MN}{}_{P}+\delta^{N}_{P}T_{Q}{}^{QM}-\delta^{M}_{P}T_{Q}{}^{QN}\rt).
\eal
Finally, the torsion scalar, which serves as a measure of torsion associated with a specific spacetime, is defined as
\bal
T={S_P}^{MN}{T^{P}}_{MN}.
\eal
The action of the 5D $f(T)$ gravity is expressed as
\bal
S=-\fc{M_*^3}{4}\int{}d^5x e f(T)+\int{}d^5x e \mc{L}_\tx{M},
\label{Main_Action}
\eal
where $M_*$ denotes the 5D fundamental gravity scale, $e=|{e^A}_M|=\sqrt{-|g_{MN}|}$, $f(T)$ is a function of the torsion scalar $T$, and $\mc{L}_\tx{M}$ is the Lagrangian for the bulk scalar field:  
\bal
\mc{L}_{\text{M}}=-\frac{1}{2}\pt^M\Phi\pt_M\Phi-V(\Phi),
\eal
which provides the material necessary to construct the brane configuration.

The field equation is obtained by varying the action \eqref{Main_Action} with respect to the vielbein and the scalar respectively, yielding
\begin{subequations}
\label{EOM}
\bal
e^{-1} f_T g_{N P} \pt_Q \lt(e S_M{ }^{P Q}\rt)+f_{T T} S_{M N}{ }^Q \pt_Q T 
-f_T \Gamma^P{ }_{Q M} S_{P N}{ }^Q+\frac{1}{4} g_{M N} f(T)&=\mc{T}_{M N}, \label{EOM_1}\\
e^{-1} \pt^K \lt(e \pt_K \Phi \rt)&=V_\Phi,\label{EOM_2}
\eal
\end{subequations}
where $f_T \equiv \fc{df(T)}{dT}$, $f_{TT} \equiv \fc{d^2f(T)}{dT^2}$, $V_\Phi \equiv \fc{dV(\phi)}{d\Phi}$, and $\mc{T}_{M N}$ is the energy-momentum tensor of the scalar field, given by
\bal
\mc{T}_{\tx{MN}}={\pt_M \Phi}{\pt_N \Phi}-\fc{1}{2} g_{MN} {\pt^P \Phi } {\pt_P \Phi }-g_{MN} V. 
\eal

To investigate the braneworld scenario, the most general 5D line element that preserves 4D Poincar\'e invariance is given by
\bal
ds^2=a^2(y)\eta_{\mu\nu}dx^\mu dx^\nu+dy^2,
\label{Brane_Metric}
\eal
where $a(y)$ is the warp factor.
Correspondingly, the proper vielbein can be chosen as $e^A{}_M=\lt(a,a,a,a,1 \rt)$. Then, with the metric and vielbein ansatz, the field equations \eqref{EOM} can be expressed explicitly as
\begin{subequations}
\label{BG_EOM}
\bal
\frac{1}{4}f + 6 H^{2} f_T &= \frac{1}{2} \Phi'^{2}-V, \label{BG_EOM_1}\\
\frac{1}{4} f + \lt(\frac{3}{2} H' + 6 H^{2}\rt) f_T - 36 H^{2} H' f_{T T}&= -\frac{1}{2} \Phi'^{2}-V, \label{BG_EOM_2}\\
\Phi''+4 H \Phi' &= V_\Phi, \label{BG_EOM_3}
\eal
\end{subequations}
where $H \equiv a'(y)/a(y)$, and the prime denotes the derivative with respect to the extra dimension coordinate $y$.

The field equations can be solved analytically by assuming an expression for the warp factor.  For a polynomial form, $f(T)=T+ \alpha T^2$, an analytical solution is given by \cite{Yang2012}
\begin{subequations}
\label{Solution_I}
\bal
a(y) &=  {\tx{sech}^b\lt(k y\rt)},\label{Solution_WF1}\\
\Phi (y) &= \sqrt{\frac{3 b}{2}} \bigg[i E(i k y,u)-i F(i k y,u)+\sqrt{1+u \sinh ^2(k y)} \tanh (k y) \bigg],\label{Solution_phi1}\\
V( \Phi(y)) &= \frac{3 b k^2}{8} \Big[ \lt(2-u+u \cosh (2 k y)\rt)\tx{sech}^4(k y)+288 \alpha  b^3 k^2 \tanh ^4(k y)-8 b \tanh ^2(k y)\Big],\label{Solution_V1}
\eal
\end{subequations}
where $b$ and $k$ are positive constants, $u\equiv 1-72 \alpha  b^2 k^2$ with the constants $\alpha b^2 k^2 \leq {1}/{72}$ to ensure that the scalar field $\Phi$ is real, and the functions $E$ and $F$ denote the elliptic integrals of the first and second kinds, respectively. 

\begin{figure}[t]
\begin{center}
\includegraphics[width=7cm,height=5cm]{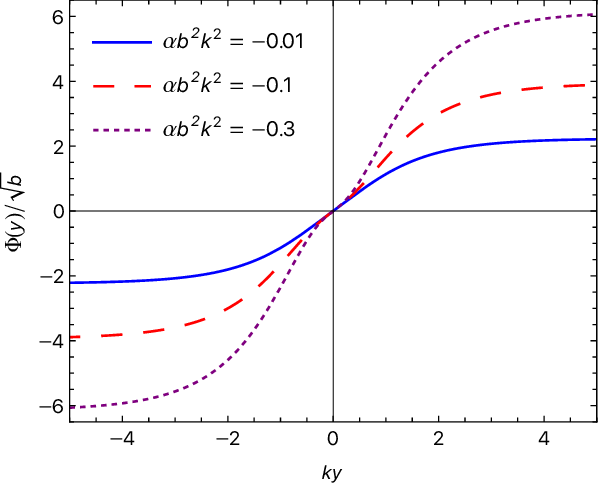}
\caption{Profiles of the scalar field $\Phi(y)$ for different parameter choices.}
\label{Scalar_profile}
\end{center}
\end{figure}

The solution describes a thick brane model with a domain wall configuration, which is a type of topological soliton. In this configuration, the scalar field non-trivially maps the boundaries of the extra dimension into two distinct scalar vacua. As shown in Fig.~\ref{Scalar_profile}, when ${1}/{72}\geq \alpha b^2 k^2 \geq -{1}/{72}$, the scalar field exhibits a kink profile. However, as $\alpha b^2 k^2<-{1}/{72}$, the scalar field transitions to a double-kink profile, and the brane exhibits an inner structure.

Particularly, for $b=1$ and $\alpha=\fc{1}{72k^2}$, the solution recovers the one derived in Ref.~\cite{Menezes2014} by using the first-order formalism\footnote{Note that we have adopted the assumption $W(\Phi)=\sqrt{6}k\Phi$ here instead of the original assumption $W(\Phi)=2\Phi$ \cite{Menezes2014} to ensure that the quantities have the appropriate dimensions.}, which reads
\begin{subequations}
\bal
a(y) &= {\text{sech}\lt(k y\rt)},\label{Solution_WF2}\\
 \Phi(y) &= \sqrt{\fc{3}{2}}\tanh \lt(k y\rt),\label{Solution_phi2}\\
V(\Phi) &= k^2 \lt[\frac{3}{4}- \Phi^2 \lt(3-\Phi^2\rt)\rt].\label{Solution_V2}
\eal
\label{Solution_II}
\end{subequations}

\section{Linear perturbations in $f(T)$ braneworlds}\label{Sec_Perturbation}

The local Lorentz violation in $f(T)$ gravity becomes more evident when the field equation \eqref{EOM_1} is reformulated as a modified Einstein equation \cite{Li2011d}, i.e.,
\bal
f_T G_{MN}+\fc{1}{2}g_{MN}\lt(f-f_T T\rt)+f_{TT}S_{MNK}\nabla^K T=\mc{T}_{MN},
\label{EOM_1_MEF}
\eal 
where $G_{MN}$ is the Einstein tensor. When $f(T) = T$, the equation recovers the Einstein equation, which is locally Lorentz invariant. However, for $f(T) \neq T$, the nonvanishing last two terms are not locally Lorentz invariant. Consequently, the broken gauge freedom in the tangent frame introduces additional DOFs in the vielbein. 

For the 4D GR, there are a total of 10 DOFs in the metric. In a spatially flat FLRW cosmology background, they can be decomposed into 4 scalars (4 DOFs), 2 transverse vectors (4 DOFs), and a symmetric transverse-traceless (TT)  tensor (2 DOFs). Since the scalar, vector, and tensor modes correspond to irreducible representations of the SO(3) group of spatial rotations, this decomposition guarantees that they do not mix in linearized theory \cite{Szapudi2012}.

However, for the 4D $f(T)$ gravity, there are additional 6 DOFs, corresponding to a vector (3 DOFs) and an antisymmetric tensor (3 DOFs). In 3D space, the antisymmetric tensor is dual to a pseudovector, i.e., $\mc{B}_{ij}=\epsilon_{ijk}B^k$ with $i,j,k=1,2,3$ corresponding to the indices of spatial coordinates. Therefore, as pointed out in Ref.~\cite{Izumi2013}, the vector and pseudovector can ultimately be decomposed into a scalar (1 DOF), a transverse vector (2 DOFs), a pseudoscalar (1 DOF), and a transverse pseudovector (2 DOFs). 

For the 5D $f(T)$ gravity, there are 10 additional DOFs in the vielbein. Correspondingly, they can be  decomposed into a 4D vector (4 DOFs) and a 4D antisymmetric tensor (6 DOFs). Due to $\mc{B}_{\mu\nu}=\epsilon_{\mu\nu\alpha\beta}B^{\alpha\beta}$ in 4D spacetime, the antisymmetric tensor is self-dual and can be equivalently expressed as an antisymmetric pseudotensor. Thus, the 10 additional DOFs can ultimately be decomposed into a scalar ($\alpha$, 1 DOF), a transverse vector ($\beta_\mu$, 3 DOFs), and an antisymmetric pseudotensor ($B_{\mu\nu}$, 6 DOFs). Consequently, the full perturbed vielbein can be expressed as 
\begin{subequations}
\label{Full_Per_vielbein}
\bal
e^a{}_\mu &=a \lt[\delta ^a{}_{\mu }(1-\varphi )+\pt^a \pt_\mu E+\pt_\mu C^a +h^a{}_{\mu}+\epsilon^a{}_{\mu \alpha \beta}B^{\alpha \beta }  \rt],\\
e^5{}_\mu &=a \lt({\pt_\mu F}+G_{\mu }+\pt_\mu \alpha+\beta _{\mu } \rt),\\
e^a{}_5 &=-\lt(\pt ^a\alpha + \beta ^a \rt),\\
e^5{}_5 &=1+\psi,
\eal
\end{subequations}
where the transverse vectors satisfy $\eta^{\mu\nu}\pt_\nu C_\mu=\eta^{\mu\nu}\pt_\nu G_\mu=\eta^{\mu\nu}\pt_\nu \beta_\mu=0$ and the transverse-traceless tensor satisfies $\eta^{\mu\lambda}\pt_\lambda h_{\mu\nu}=\eta^{\mu\lambda}h_{\mu\nu}=0$. 

With the full perturbed vielbein, the perturbed metric at linear order is expressed as
\begin{subequations}
\bal
g_{\mu\nu}&=a^2\lt[(1-2\varphi)\eta_{\mu\nu}+2\pt_\mu\pt_\nu E +\pt_{(\mu} C_{\nu)}+2h_{\mu\nu}\rt], \\
g_{\mu5}&=a\lt(\pt_\mu F+G_\mu \rt), \\
g_{55}&=1+2\psi.
\eal
\end{subequations}
This is the standard perturbed metric in GR without additional DOFs ($\alpha$, $\beta_\mu$, and $B_{\mu\nu}$) associated with Lorentz breaking. Moreover, the corresponding perturbed scalar field is expressed as $\Phi = \bar{\Phi} + \phi$, where $\bar{\Phi}$ represents the background scalar.

Although the $f(T)$ gravity theory is formulated using orthogonal vielbein fields, as shown in the field equation \eqref{EOM_1_MEF}, it still utilizes covariant tensors to describe the dynamics of spacetime. Consequently, similar to GR, the action of the $f(T)$ gravity remains invariant under the general coordinate transformations $x^M\to {x'}^M=x^M+\xi^M(x)$. Under the transformations, the vielbein transforms as
\bal
\delta e^A{}_M=-\bar{e}^A{}_{N}{\pt_M}{\xi^N}-\xi^N \pt_N \bar{e}^A{}_M,
\eal
where $\bar{e}^A{}_{M}$ denotes the background vielbein. Furthermore, by decomposing the 4D component of $\xi^M$ as $\xi_\mu = a^{2}(\partial_\mu \xi^{\|} + \xi^\perp_\mu)$ with $\xi^\perp_\mu$ a transverse vector and $\xi^{\|}$ a longitudinal scalar, the gauge transformations of the perturbed quantities can be derived as
\bal
&\Delta \phi = -\bar\Phi' \xi^5, ~
\Delta  \psi = -\xi^{5 \prime }, ~
\Delta\varphi = H \xi^5,  ~
\Delta E= -\xi^{\|}, ~ \Delta\alpha = a \lt({\xi^{\|}}' + 2H \xi^{\|} \rt), ~
\Delta F = -a\lt({\xi^{\|}}'+2H\xi^{\|} \rt)-{a^{-1}}{\xi^5},~\nn\\
&\Delta \beta_\mu = a \lt({\xi^{\perp}_{\mu}}'+2H\xi^{\perp}_{\mu}\rt), ~
\Delta G_{\mu } = -a\lt({\xi^{\perp}_{\mu}}'+2H\xi^{\perp}_{\mu}\rt), ~ 
\Delta C_{\mu } = -\xi^{\perp}_{\mu}, ~
\Delta h_{\mu\nu} = 0, ~
\Delta B_{\mu\nu} = 0. 
\label{Gauge_Transformation}
\eal
It is clear that the tensor mode $h_{\mu\nu}$ and the pseudotensor  mode $B_{\mu\nu}$ are gauge invariant. However, 2 scalar modes and 1 vector mode can be directly gauged out by appropriately choosing $\xi^5$, $\xi^{\|}$ and $\xi^\perp_\mu$. 

The scalar, vector, antisymmetric pseudotensor, and symmetric tensor modes do not mix with each other under the 4D Lorentz transformation, allowing us to treat them separately in linearized theory.

\subsection{Scalar perturbations}

From the gauge transformations \eqref{Gauge_Transformation}, we can properly select the  scalars $\xi^{\|}$ and $\xi^5$ to eliminate two scalar modes in perturbation theory. Specifically, we first gauge out the scalar mode $E$ in the perturbed vielbein \eqref{Full_Per_vielbein} by fixing $\xi^{\|}$. Consequently, the vielbein with scalar perturbations reads 
\bal
e^A{}_M=\lt(
\begin{array}{cc}
 a\delta ^a{}_{\mu }(1-\varphi ) & -\pt^a\alpha  \\
a \pt_\mu \mc F & 1+\psi \\
\end{array}
\rt),
\label{S_Per_Vielbein}
\eal  
where $\mc F \equiv F+\alpha$.

With the perturbed vielbein \eqref{S_Per_Vielbein}, the nonvanishing components of the torsion tensor, up to second order in perturbations, are given by
\bal
T^5{}_{\mu \nu }&= - a{\pt_{[\mu}}{\mc F}{\pt_{\nu]}}{\varphi},\\
T^5{}_{5 \mu }&=-T^5{}_{\mu 5}=a (1\!-\!\psi ) {\pt_\mu }{ \mc F'}-(1-\psi){\pt_\mu }{\psi} +a {\varphi}' {\pt_\mu} {\mc F}-{\pt_\nu}{\mc F}\pt ^{\nu }{\pt_\mu }{\alpha },\\
T^{\rho }{}_{5 \mu }&=-T^{\rho }{}_{\mu 5}=\delta^{\rho}{}_{\mu }H - \delta^{\rho}{}_{\mu }(1+\varphi) \varphi '+ a^{-1}\pt ^{\rho }{\pt_\mu }{\alpha }-a^{-1}\pt^{\rho }\alpha {\pt_\mu }{\psi }+a^{-1}\varphi \pt^{\rho }{\pt_\mu }{ \alpha }+\pt^{\rho }\alpha {\pt_\mu }{ \mc F'},\\
T^{\rho }{}_{\mu \nu }&=(1+\varphi)\delta^{\rho }{}_{[\mu}\pt_{\nu]}{\varphi},
\eal
where the indices are raised and lowered by the 4D Minkowski metric $\eta_{\mu\nu}$. Correspondingly, the perturbed torsion scalar reads
\bal
T&=-12 {H}^2+6H\lt(4 H \psi+4 \varphi '-{a}^{-1}\pt^{\gamma }{\pt_\gamma }{\alpha }\rt)-36 H^2 \psi ^2+24 H \varphi  \varphi '-12 {\varphi '}^2-48 H \psi  \varphi '-{a^{-2}}\big(6\pt^{\gamma }\varphi {\pt_ \gamma }{\varphi }\nn\\
&-6\pt^{\gamma }\varphi{\pt_ \gamma }{ \psi }+\pt^{\gamma }{\pt_\gamma }{\alpha }\pt^{\rho }{\pt_\rho }{ \alpha }-\pt_{\gamma }\pt^{\rho }\alpha \pt^{\gamma }{\pt_\rho }{ \alpha }\big)+{a^{-1}}H\big(6\pt^{\gamma }\mc F{\pt_\gamma }{ \psi }-18\pt^{\gamma }\mc F{\pt_\gamma }{\varphi }+18\pt^{\gamma }\varphi{\pt_ \gamma } {\alpha }-6\varphi \pt^{\gamma }{\pt_\gamma }{\alpha }\nn\\
&+12\psi \pt^{\gamma }{\pt_\gamma }{\alpha }\big)+6{a^{-1}}\big(\varphi '\pt^{\gamma }{\partial_ \gamma }{ \alpha }-\pt^{\gamma }\varphi{\pt_\gamma } { \mc F'}\big)-6H\pt^{\gamma }\mc F{\pt_\gamma }{ \mc F'}-12H^2\pt^{\gamma }\mc F{\pt_\gamma }{\mc F}+24H^2\pt^{\gamma }\mc F{\pt_\gamma }{\alpha }.
\eal

With the action \eqref{Main_Action} and the background equation \eqref{BG_EOM}, after integrating by parts and dropping the boundary terms, the quadratic action for scalar perturbations is obtained as
\bal
S^{(2)}_\tx{S}&=M_*^3 \int{d^4xdy}\bigg[
  -\fc{9}{2} a^2 H^2 f_{{TT}} \pt^{\lambda }{\pt_\lambda }{\alpha }\pt^{\rho }{\pt_\rho }{\alpha}-\fc{a^2}{2} \pt^{\lambda }\phi {\pt_\lambda}\phi+\fc{3}{2} a^2 f_T \pt^{\lambda }\varphi {\pt_\lambda }{\varphi }+\fc{3}{4} a^4 \lt(4 H^2+H'\rt) \lt(f_T-24 H^2 f_{{TT}}\rt)\psi^2
 \nn\\
& 
 +3 a^4\lt(f_T-24 H^2 f_{{TT}}\rt) {\varphi '}^2-\fc{1}{2} a^4 V_{\Phi \Phi} \phi^2 
 -\fc{1}{2} a^4 {\phi'}^2 - \fc{3}{2}a^2f_T\pt^{\lambda }\varphi {\pt_\lambda }{\psi }
 + \fc{3}{2}a^3 H f_T{ \psi }\pt^{\lambda }{\pt_\lambda }\mc F + \fc{3}{2} a^3 f_T\varphi'\pt^{\lambda }{\pt_\lambda }{\mc F}\nn\\
 &- 36a^3 H H' f_{TT}\varphi \pt^{\lambda }{\pt_\lambda }{\mc F}
 - \frac{3}{2} a^3  \lt(f_T - 24 H^2 f_{{TT}}\rt) \varphi'\pt^{\lambda }{\pt_\lambda }{\alpha }+\fc{3}{2} a^3 H \lt(f_T-24 H^2 f_{{TT}}\rt)\pt^{\lambda }\psi {\pt_\lambda }{\alpha }
 -a^3 \bar\Phi' \phi \pt^{\lambda }{\pt_\lambda }{\mc F}
 \nn\\
&+a^3 \bar\Phi'\phi \pt^{\lambda }{\pt_\lambda }{\alpha}+6 a^4 H\lt(f_T-24 H^2 f_{{TT}}\rt) \varphi' \psi  
- a^4 V_{\Phi} \phi \psi
 + 4 a^4 V_{\Phi}   \varphi \phi + 4 a^4 \Phi' \phi' \varphi  
 +a^4 \bar\Phi'\phi'\psi  \bigg].
 \label{Action_Scalar}
\eal

Here, we still have one remaining gauge freedom. Due to $\Delta\delta^{(1)}T = -T' \xi^5$ under general coordinate transformations, we adopt the torsion gauge instead of the usual approach of eliminating a specific scalar mode. This choice necessitates that the linear perturbation of the torsion scalar $\delta^{(1)}T$ vanishes. This is akin to the choice of curvature gauge in the analysis of scalar perturbations in $f(R)$ gravity \cite{Zhong2017,Zhong2018}. Thus, the torsion gauge provides a constraint
\bal
4 H \psi+4 \varphi '-{a}^{-1}\partial ^{\gamma }{\partial_\gamma }{\alpha }=0.
\label{Cons_S_T}
\eal

Furthermore, by varying the quadratic action \eqref{Action_Scalar} with respect to the modes $\mc F$, $\alpha$, and $\psi$ respectively, we derive additional constraint equations as
\bal
\frac{3}{2} f_T \lt(H \psi +\varphi '\rt)-36 H f_{{TT}} H'\varphi - \bar\Phi' \phi = 0,& \label{Cons_S_F} \\
\frac{3}{2} a \lt(f_T-24 H^2 f_{TT}\rt)\varphi'+ \frac{3}{2} a H \lt(f_T-24 H^2 f_{TT}\rt) \psi 
+9H^2f_{\text{TT}}\pt^{\lambda}\pt_\lambda \alpha  - a \bar\Phi' \phi  = 0, & \label{Cons_S_alpha}\\
\frac{3}{2} a^2  H' \lt(f_T-24 H^2 f_{{TT}}\rt)\psi+\fc{3}{2}f_T\pt ^{\lambda }{\pt_\lambda }{ \varphi }
+\fc{3}{2}aHf_T\pt^{\lambda }{\pt_\lambda }{\mc F}-a^2 V_\Phi \phi+a^2 \bar\Phi' \phi ' =0. &\label{Cons_S_psi}
\eal

By employing the relation \eqref{Cons_S_T}, the constraint \eqref{Cons_S_alpha} can be further simplified to
\bal
\frac{3}{2} f_T \left(H \psi +\varphi '\right)-\bar\Phi'\phi =0.
\label{Cons_S_alpha_2}
\eal
Combining this with the constraint \eqref{Cons_S_F}, we obtain
\bal
\varphi=0.
\eal
Then, Eq.~\eqref{Cons_S_alpha_2} simply yields
\bal
\psi =\frac{2 \bar\Phi'}{3 H f_T}\phi.
\eal
With the expressions of $\varphi$ and $\psi$, we can work out $\pt^\lambda \pt_\lambda \alpha$ and $\pt^\lambda \pt_\lambda F$ from Eqs.~\eqref{Cons_S_T} and \eqref{Cons_S_psi}, yielding
\bal
\pt^{\lambda }{\pt_\lambda }{\alpha }&=\fc{8 a \bar\Phi' }{3 f_T}\phi,\\
 \pt^{\lambda }{\pt_\lambda }{\mc F}&=\fc{2 a}{3 H^2 f_T^2}\Big[ f_TH V_\Phi \phi-f_T \bar\Phi' \lt(H' \phi  +H \phi' \rt)+24  H^2 f_{TT} H' \bar\Phi' \phi\Big].
\eal
Further, by substituting the results into the quadratic action \eqref{Action_Scalar} and simplifying through some simple algebra, we  arrive at
\bal
\!S^{(2)}_\tx{S}&=-\fc{M_*^3}{2}\int{d^4xdy}a^2\bigg[\pt^{\lambda} \phi {\pt_\lambda }{\phi}+ a^2  {\phi'}^2+a^2\bigg(V_{\Phi\Phi}+\frac{16 \bar{\Phi}^{'2}}{3 f_T}+\frac{8 {\bar\Phi}^{'4}}{9 H^2 f_T^2}+\frac{8 \bar\Phi' \bar\Phi''}{3 H f_T}\bigg)\phi ^2\bigg].
\eal
Thus, it is evident that there is only one scalar propagating DOF. Specifically, the correct sign of the kinetic term ensures that there is no ghost scalar mode in the linearized theory. 

In order to study the mass spectrum of the scalar mode, it is convenient to rewrite the action in the conformal extra dimension coordinate $z$, which relates to the physical extra dimension coordinate as $dy = a dz$, yielding
\bal
S^{(2)}_\tx{S}=-\frac{M_*^3}{2}\int{d^4x dz}a^3\lt[\pt^{\lambda}\phi {\pt_\lambda}{\phi}+\dot{\phi}^2+K(z)\phi ^2 \rt],
\label{Action_S_Fin}
\eal
where the dot denotes the derivative with respect to the conformal coordinate $z$, and
\bal
K(z)=a^2 V_{\Phi\Phi }+\frac{8 \dot{\bar\Phi }^2}{3 f_T}+\fc{8 \dot{\bar\Phi }^4}{9 H^2 f_T^2}+\frac{8\dot{\bar\Phi }\ddot{\bar\Phi }}{3 H f_T},
\label{Expression_Hz}
\eal
with $H\equiv {\dot a(z)}/{a(z)}$ in $z$ coordinate.

By varying the action \eqref{Action_S_Fin} with respect to $\phi$, we obtain the equation of motion
\bal
\Box^{(4)}{\phi }+\ddot{\phi}+3H\dot{\phi}-K\phi =0,
\label{EOM_Scalar}
\eal
where $\Box^{(4)}\equiv\eta^{\alpha\beta}\pt_{\alpha }{\pt_\beta }$ is the 4D d'Alembert operator. 

With a KK decomposition in the form $\phi(x,z) =\hat \phi (x)\chi_\tx{S}(z)a^{-\frac{3}{2}}(z)$, the equation of motion \eqref{EOM_Scalar} can be further decomposed into a Klein-Gordon equation $\Box^{(4)}\hat\phi=m^2\hat\phi$ and a Schr\"odinger-like equation
\bal
-\ddot{\chi}_\tx{S}(z)+U_\tx{S}(z)\chi_\tx{S}(z) ={m^2}\chi_\tx{S}(z),
\eal
where $m$ is the mass of scalar KK modes, and the effective potential reads
\bal
U_\tx{S}(z)=\frac{9}{4}H^2+\frac{3}{2} \dot H+K.
\label{Effective_Potential}
\eal

\begin{figure}[t]
\begin{center}
\subfigure[~Different $\alpha$]  {\label{Potential_alpha}
\includegraphics[width=7cm,height=5cm]{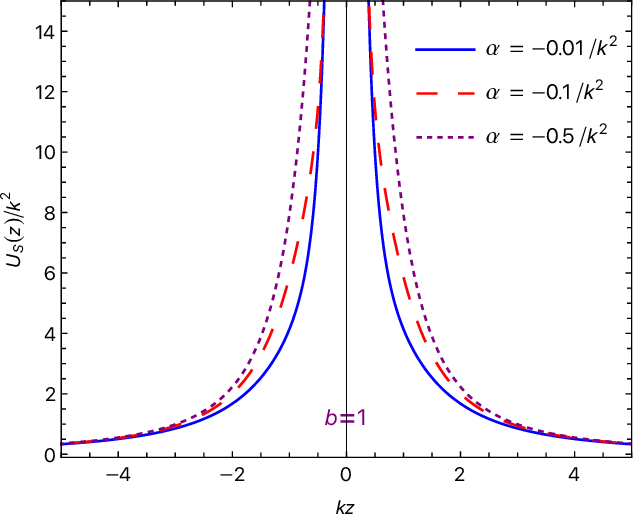}}
\qquad
\subfigure[~Different $b$]  {\label{Potential_b}
\includegraphics[width=7cm,height=5cm]{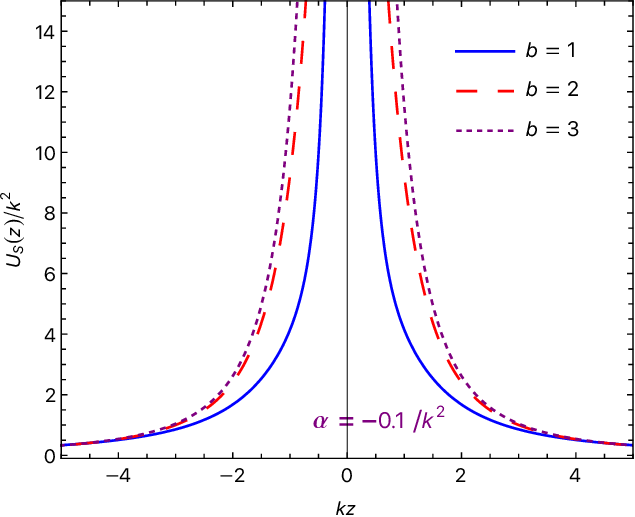}}
\caption{Profiles of the effective potential $U_\tx{S}(z)$ for different parameter choices.}
\label{Effective_Potentials}
\end{center}
\end{figure} 

The theory would be free from tachyonic instability if the masses of all scalar KK modes are non-negative. Since the mass spectrum is entirely governed by the effective potential, it is essential to analyze the behavior of this potential. Given the complexity of its expression, the effective potentials for the solution described in \eqref{Solution_I} with varying values of the parameters $\alpha$ and $b$ are illustrated in Fig.~\ref{Effective_Potentials}.

It is evident that the potential presents a barrier that diverges at the origin and asymptotically approaches zero as $z\to \pm \infty$. Consequently, all scalar KK modes correspond to free states with a mass spectrum of $m^2>0$, ensuring that the scalar perturbations are robust and free from tachyonic instability. Additionally, the infinite barrier results in the complete reflection of any scalar KK mode back to infinity. As a result, the scalar modes are decoupled and do not contribute to the low-energy effective theory on the brane.

\subsection{Vector perturbations}

For the vector perturbations, we have the gauge freedom $\xi^\perp_\mu$ to eliminate the mode $C_\mu$ in the perturbed vielbein, yielding
\bal
e^A{}_M=\lt(
\begin{array}{cc}
 a\delta^a{}_{\mu } & -\beta^a  \\
a\lt(G_{\mu}+\beta_\mu \rt) & 1 \\
\end{array}
\rt).
\label{V_Per_Vielbein}
\eal  
Using the perturbed vielbein, we can derive the nonvanishing components of the torsion tensor up to second order in perturbations:
\bal
T^5{}_{\mu \nu} &= a\lt( {\pt_{[\mu} }{ G_{\nu]}}+{\pt_{[\mu}}{\beta _{\nu] }}\rt), \\
T^5{}_{5 \mu} &= -T^5{}_{\mu 5} =a \lt(G_{\mu}'+\beta_{\mu}'\rt)-\lt(G^{\lambda}+\beta^{\lambda}\rt){\pt_\mu }{\beta_{\lambda}}, \\ 
T^\rho{}_{5 \mu} &= -T^\rho{}_{\mu 5} = H \delta^{\rho}{}_{\mu}+{a^{-1}}{\pt_\mu }{{\alpha^{\rho}}}+\alpha^{\rho} \lt(G_{\mu}'+\alpha_{\mu}'\rt), \\
T^\rho{}_{\mu \nu} &= \alpha^{\rho} \lt({\pt_{[\mu} }{G_{\nu]}}+{\pt_{[\mu} }{ \alpha_{\nu]}}\rt).
\eal
Then, the perturbed torsion scalar is given by
\bal
T&=-12 H^2-6{a^{-1}}H\pt^{\lambda }\beta_{\lambda }-12 H^2 \lt(G^{\lambda } G_{\lambda }-\beta^{\lambda } \beta_{\lambda }\rt)+\frac{1}{2 a^2}\lt(\pt^{\lambda}G^{\rho }{\pt_\lambda }{ G_{\rho }}-\pt^{\lambda }G_{\rho }\pt^{\rho }G_{\lambda }-2\pt^{\lambda }\beta _{\lambda }\pt^{\rho }\beta _{\rho }+2\pt^{\lambda }\beta_{\rho }\pt^{\rho }\beta_{\lambda }\rt)\nn\\
& -6 H \lt(G^{\lambda } G_{\lambda }'+G^{\lambda }\beta_{\lambda }'+G_{\lambda }' \beta^{\lambda } +\beta^{\lambda } \beta_{\lambda }'\rt).
\eal

From the action \eqref{Main_Action}, after integrating by parts and dropping the boundary terms, the quadratic action for vector perturbations is given by
\bal
S^{(2)}_\tx{V}&=-\frac{M_*^3}{4} \int{d^4xdy} a^4\bigg[\fc{1}{2}a^{-2}f_T\partial ^{\lambda }G^{\rho }{\partial_ \lambda }{G_{\rho }}+\Big(3 H' f_T-72 H^2  H' f_{TT}+2 {\Phi'}^2 \Big) G^{\lambda } G_{\lambda } +\big(f+4 V +3 H'f_T \nn\\
&  + 24  H^2f_T  -72 H^2 f_{\text{TT}} H'\big)\alpha ^{\lambda } \alpha _{\lambda }+\Big(f+4 V+2 {\Phi'}^2+6H'f_T+24 H^2f_T-144 H^2 f_{TT} H'\Big) G^{\lambda }\alpha_{\lambda } \bigg].
\eal
By using the background equations \eqref{BG_EOM}, the action can be further simplified as
\bal
S^{(2)}_\tx{V}&=-\frac{M_*^3}{8} \int{d^4xdy} a^2f_T\partial ^{\lambda }G^{\rho }{\partial_ \lambda }{G_{\rho }}.
\eal
It is clear that the sign of the kinetic term is entirely determined by $f_T$. When $f_T$ is positive, the vector mode is ghost-free. Conversely, when  $f_T$ is negative, it corresponds to a ghost mode.

For the solution \eqref{Solution_I}, the function $f_T$ is given by
\bal
f_T(y)=1+2\alpha T=1-24\alpha b^2k^2 \tanh ^2(k y),
\eal
which is always positive. We can define $\mc G={\sqrt{f_T}}G/{2}$ to achieve a canonically normalized form in the conformal coordinate $z$:
\bal
S^{(2)}_\tx{V}&=-\frac{M_*^3}{2} \int{d^4xdz} a^3\pt^{\lambda }\mc{G}^{\rho }{\pt_\lambda }{\mc{G}_{\rho }}.
\eal
In this case, the vector mode is massless and robust.

\subsection{Pseudotensor perturbation}

For the pseudotensor perturbation, the perturbed vielbein reads
\bal
e^A{}_M=\lt(
\begin{array}{cc}
 a\lt(\delta^a{}_{\mu }+\epsilon^a{}_{\mu\alpha\beta}B^{\alpha\beta}\rt) & 0  \\
0 & 1 \\
\end{array}
\rt).
\label{PT_Per_Vielbein}
\eal
After some straightforward algebra, the nonvanishing components of the perturbed torsion tensor are expressed as
\bal
T^{\rho }{}_{5 \mu } &= H \delta ^{\rho }{}_{\mu }+{\epsilon ^\rho}_{\mu \tau \kappa }{B^{\tau \kappa }}'+
{{\epsilon}^{ \alpha\rho } }_{\sigma \omega } {\epsilon_{\alpha \mu \tau \kappa }} B^{\sigma \omega }{B^{\tau \kappa }}',\\
T^{\rho }{}_{\mu \nu } \!&=\! \epsilon ^{\rho }{}_{\nu \tau \kappa } {\partial _\mu }{B^{\tau \kappa }}
-\epsilon ^{\rho }{}_{\mu \tau \kappa } {\pt_\nu }{B^{\tau \kappa }}+{{\epsilon} ^{\alpha\rho }}_{\sigma \omega  } \epsilon_{\alpha \nu \tau \kappa }  B^{\sigma \omega }{\partial _\mu }{B^{\tau \kappa }}
- {{\epsilon} ^{\alpha\rho }}_{\sigma \omega  } \epsilon_{\alpha \mu \tau \kappa }  B^{\sigma \omega  }{\partial _\nu }{B^{\tau \kappa }}.
\eal
Then, the perturbed torsion scalar reads 
\bal
T&=-12 H^2
-6 H \epsilon^{\alpha \beta}{}_{\sigma \omega} {{\epsilon}_{\alpha \beta  \tau \kappa  }}B^{\sigma \omega }{{B^{\tau \kappa}}}^{ \prime }
+a^{-2}\epsilon ^{\rho }{}_{\beta \sigma \omega } \epsilon ^{\alpha \beta }{}_{\tau \kappa }  \lt({\pt_\alpha  B^{\sigma \omega }} {\pt_\rho B^{\tau \kappa }}- {\pt_\alpha B^{\tau \kappa }} {\pt_\rho B^{\sigma \omega }}\rt).
\eal

From the action \eqref{Main_Action}, after integrating by parts and dropping the boundary terms, the quadratic action for pseudotensor perturbation is given by
\bal
S^{(2)}_\tx{PT}&=-\frac{M_*^3}{2} \int{d^4xdy}a^4 \epsilon ^{\alpha \beta }{}_{\sigma \omega }  \epsilon_{\alpha \beta \tau \kappa }\bigg[\lt(\fc{f}{4}+\fc{{\Phi'}^2}{2}+V\rt)B^{\sigma \omega } B^{\tau \kappa }-3 H f_T B^{\sigma \omega }  {B^{\tau \kappa }}'\bigg].
\eal
By using the background equation \eqref{BG_EOM_2} and the identity $\epsilon ^{\alpha \beta }{}_{\sigma \omega }  \epsilon_{\alpha \beta \tau \kappa }=2\lt(\eta_{\kappa\sigma} \eta_{\tau\omega}-\eta_{\tau\sigma} \eta_{\kappa\omega}\rt)$, the action can be further simplified as
\bal
S^{(2)}_\tx{PT}&=2{M_*^3} \int\!\!{d^4xdy}a^4 
\bigg[ \bigg(36 H^2 H'f_{TT} -\frac{3 }{2}H' f_T - 6 H^2 f_T \bigg)B_{\alpha\beta } B^{\alpha\beta}-3 H f_T B_{\alpha\beta }  {B^{\alpha\beta }}' \bigg].
\eal
After integrating the last term by parts and neglecting the boundary terms, it exactly cancels the first term in the action. Therefore, the vanishing Lagrangian $\mc{L}^{(2)}_\tx{PT}=0$ implies that, up to second order in perturbations, there is no propagating pseudotensor mode in the theory.

\subsection{Tensor perturbation}

For the tensor perturbation, the perturbed vielbein reads
\bal
e^A{}_M=\lt(
\begin{array}{cc}
 a\lt(\delta^a{}_{\mu }+h^a{}_{\mu}\rt) & 0  \\
0 & 1 \\
\end{array}
\rt).
\label{T_Per_Vielbein}
\eal
Consequently, the nonvanishing components of the torsion tensor, up to second order in perturbations, are calculated as follows:
\bal
T^{\rho }{}_{5 \mu }&=-T^{\rho }{}_{\mu 5}=H \delta ^{\rho }{}_{\mu }+h^{\rho }{}_{\mu }'-h^{\rho }{}_{\lambda } h^{\lambda }{}_{\mu }',\\
T^{\rho }{}_{\mu \nu }&={\pt_{[\mu} }{h^\rho{}_{\nu] }} - h^{\rho }{}_\lambda {\pt_{[\mu} }{ h^\lambda{}_{\nu]}}. 
\eal
Moreover, the perturbed torsion scalar is obtained as
\bal
T&=-12 H^2+{a^{-2}}(\pt^{\lambda }h^{\alpha\beta } {\pt_\lambda }{ h_{\alpha\beta}}-{\pt^\lambda } { h^{\alpha \beta }}{\pt_\alpha }{ h_{\lambda\beta }})+{h^{\alpha \beta }}' h_{\alpha \beta }'+6 H h^{\alpha\beta } h_{\alpha\beta }'.
\eal

With the action \eqref{Main_Action} and the background equation \eqref{BG_EOM_2}, after integrating by parts and neglecting the boundary terms, the quadratic action for tensor perturbations is obtained as
\bal
S^{(2)}_\tx{T}=&-\fc{M_*^3}{4}\int {dx}^4dy a^2 f_T \lt( \pt^\lambda h^{\alpha\beta }\pt_\lambda h_{\alpha\beta } +a^2 {h^{\alpha\beta }}' h'_{\alpha\beta }\rt).
\eal
Similar to the vector perturbations, the sign of the kinetic term is determined by $f_T$. Therefore, the tensor perturbation is ghost-free for the solution \eqref{Solution_I}.

By varying the action with respect to $h^{\alpha\beta}$, one obtains the equation of motion for the tensor mode as  
\bal
 f_T\lt(\gamma_{\alpha\beta}''+4 H \gamma_{\alpha\beta}+a^{-2}\Box^{(4)}\gamma_{\alpha\beta}\rt)-24 H H' f_{{TT}} \gamma_{\alpha\beta}'=0,
\label{EOM_Tensor}
\eal
which was first presented in Ref.~\cite{Guo2016}. In the conformal coordinate $z$, the equation can be rewritten as
\bal
\lt(\pt_z^2+\mc K \pt_z+\Box^{(4)}\rt) \gamma_{\mu \nu}=0,
\eal
where  
\bal
\mc K \equiv 3H+\fc{24f_{TT}}{a^2f_T}\lt(H^3-H \dot H\rt).
\eal
With a KK decomposition $\gamma_{\alpha\beta }  =\hat\gamma_{\alpha\beta }(x) \chi_\tx{T}(z) e^{-\int{}\fc{\mc K}{2}}dz $, the equation of motion \eqref{EOM_Tensor} can be decomposed into a Klein-Gordon equation $\Box^{(4)}\hat\gamma_{\alpha\beta }  =m^2\hat\gamma_{\alpha\beta }$ and a Schr\"odinger-like equation
\bal
-\ddot{\chi}_\tx{T}(z)+U_\tx{T}(z)\chi_\tx{T}(z) ={m^2}\chi_\tx{T}(z),
\eal
where $m$ is the mass of KK gravitons, and the effective potential is given by
\bal
U_\tx{T}(z)=\frac{\dot{\mc{K}}}{2}+\fc{\mc{K}^2}{4}.
\label{Tensor_Effective_Potential}
\eal
The Hamiltonian can be factorized as a supersymmetric quantum mechanics form
\bal
\mc{H}=-\pt_z^2+U_\tx{T}=A^\dag A=\lt(\pt_z+\fc{\mc{K}}{2}\rt)\lt(-\pt_z+\fc{\mc{K}}{2}\rt).
\eal
Given a Neumann boundary condition $\pt_z\gamma_{\alpha\beta}|_{z\to\pm\infty} = 0$, it can be shown that the eigenvalues $m^2$ are non-negative \cite{Yang2017}. Consequently, there are no tachyonic modes in the tensor perturbation, indicating that the theory is stable against the tensor perturbation.

For the tensor KK modes, it is easy to check that there is a localized massless graviton, whose wave function reads
\bal
\chi_\tx{T}=N_0e^{\int{}\fc{\mc{K}}{2}dz},
\eal
where $N_0$ is the normalization factor. The massless mode refers to the 4D graviton on the brane. As shown in Refs.~\cite{Guo2016,Yang2018}, the massless mode is localized on the brane, and all the massive KK modes are continuous free states.

\section{Conclusions}\label{Conclusions}

In this work, we investigated the full linear perturbations of the thick brane scenario in $f(T)$ gravity. Due to the local Lorentz violation in $f(T)$ gravity, the theory possesses more DOFs compared to GR. In 5D spacetime, there are a total of 25 DOFs in the perturbed vielbein, which can be decomposed into 5 scalars, 3 transverse vectors, 1 antisymmetric pseudotensor, and 1 symmetric transverse-traceless tensor. By appropriately choosing the gauge, we calculated the quadratic action for each mode and analyzed their stability.

Working in the gauge $E=\delta^{(1)}T=0$, we found that there is only one propagating DOF among the scalar perturbations. For the vector perturbations, our research indicates that there is only one massless propagating DOF, and its stability is entirely determined by the sign of $f_T$. When $f_T$ is positive, the vector mode is stable. In contrast, when $f_T$ is negative, the vector mode suffers from the ghost instability. Interestingly, we found that there is no propagating DOF in the pseudotensor perturbation. For the tensor perturbation, it represents a propagating DOF, and its stability is also determined by the sign of $f_T$, similar to the case of the vector mode. For the thick brane solution \eqref{Solution_I} with $f(T)=T+ \alpha T^2$, all the scalar, vector, pseudotensor, and tensor perturbations are stable.

As is well known, due to the relation $R=-T+2\nabla_P T_Q{}^{QP}$ \cite{Bahamonde2023}, teleparallel gravity is dynamically equivalent to GR. When $f(T)=T$, $f(T)$ gravity reduces to teleparallel gravity, and consequently, to GR. In this case, it can be directly verified that for the flat brane solutions in GR, such as the sine-Gordon solution \cite{Gremm2000a}, our results indicate the presence of continuous massive scalar modes, a massless vector mode, and both massless and continuous massive tensor modes, with all perturbations being stable. These findings are consistent with the analysis of thick brane stability in GR \cite{DeWolfe2000,Giovannini2001,Kobayashi2002,Aybat2010}. Interestingly, for flat braneworlds in $f(R)$ gravity, the scalar perturbations are stable, and the stability of the vector and tensor perturbations is similarly determined by the sign of $f_R$. When $f_R$ is positive, the vector and tensor perturbations are also stable \cite{Zhong2017,Zhong2018}. 

The propagating DOFs in the $f(T)$ braneworld consist of one scalar mode, one massless vector mode, and one tensor mode. Therefore, under the flat braneworld background \eqref{Brane_Metric}, no additional DOFs appear in the linearized theory compared to the case of GR. This conclusion is consistent with the findings from the study of linear perturbations in $f(T)$ gravity cosmology \cite{Izumi2013}, which also observed that no additional DOFs emerge at the linear level. In general, there are four additional DOFs for the 5D $f(T)$ gravity, as noted by Ref.~\cite{Li2011a}. These additional DOFs may emerge at the nonlinear level or in less symmetric braneworld backgrounds. However, this is beyond the scope of the current paper and will be left for future investigation.

\section*{ACKNOWLEDGMENTS}

This work was supported by the National Natural Science Foundation of China (Grant Nos.~12475062 and 12005174) and the Natural Science Foundation of Chongqing (Grant No.~CSTB2024NSCQ-MSX0358). KY also acknowledges the generous hospitality during a visit to the Lanzhou Center for Theoretical Physics, where part of this work was completed.


\end{document}